\documentclass{webofc}
\usepackage[varg]{txfonts}
\usepackage{multirow}

\newcommand{\erosita}{{\it eROSITA}}
\newcommand{\xrism}{{\it XRISM}}
\newcommand{\planck}{{\it Planck}}
\newcommand{\arcmin}{$^{\prime}$}
\def\xmm{XMM-{\it Newton}}

\begin{document}

\title{OLIMPO: a Balloon-Borne SZE Imager to Probe ICM Dynamics and the WHIM}

\author{\lastname{J. Sayers}\inst{1}\fnsep\thanks{\email{jack@caltech.edu}} \and
        \lastname{C. Avestruz}\inst{2} \and
        \lastname{R. Basu Thakur}\inst{3} \and
        \lastname{E. Battistelli}\inst{4} \and
        \lastname{E. Bulbul}\inst{5} \and
        \lastname{F. Cacciotti}\inst{4} \and
        \lastname{F. Columbro}\inst{4} \and
        \lastname{A. Coppolecchia}\inst{4} \and
        \lastname{S. Cray}\inst{6} \and
        \lastname{G. D'Alessandro}\inst{4} \and
        \lastname{P. de Bernardis}\inst{4} \and
        \lastname{M. De Petris}\inst{4} \and
        \lastname{S. Hanany}\inst{6} \and
        \lastname{L. Lamagna}\inst{4} \and
        \lastname{E. Lau}\inst{7} \and
        \lastname{S. Masi}\inst{4} \and
        \lastname{A. Paiella}\inst{4} \and
        \lastname{G. Pettinari}\inst{8} \and        
        \lastname{F. Piacentini}\inst{4} \and
        \lastname{E. Rapaport}\inst{1} \and
        \lastname{L. Rudnick}\inst{6} \and
        \lastname{I. Zhuravleva}\inst{9} \and
        \lastname{J. ZuHone}\inst{7}}

\institute{Caltech, 1200 E. California Blvd., Pasadena, CA 91125, USA \and
           University of Michigan, 450 Church St., Ann Arbor, MI 48109, USA \and
           NASA Jet Propulsion Laboratory, 4800 Oak Grove Dr., Pasadena, CA 91011, USA \and
           Physics Department, Sapienza University of Rome, Piazzale Aldo Moro 5, 00185 Rome, Italy \and
           Max-Planck-Institut f\"{u}r extraterrestrische Physik, Giessenbachstra{\ss}e 1, D-85748 Garching, Germany \and
           University of Minnesota, 115 Union St. SE, Minneapolis, MN 55455, USA \and
           Center for Astrophysics | Harvard \& Smithsonian, 60 Garden St., Cambridge, MA 02138, USA \and
           Instituto di Fotonica e Nanotecnologie (IFN) - Consiglio Nazionale delle Ricerche (CNR) - Via del Fosso del Cavaliere 100, 00133 Rome, Italy \and
           The University of Chicago, Chicago, IL 60637, USA}

\abstract{OLIMPO is a proposed Antarctic balloon-borne Sunyaev-Zel'dovich effect (SZE) imager to study gas dynamics associated with structure formation along with the properties of the warm-hot intergalactic medium (WHIM) residing in the connective filaments. During a 25 day flight OLIMPO will image a total of 10 $z$$\sim$0.05 galaxy clusters and 8 bridges at 145, 250, 365, and 460 GHz at an angular resolution of 1.0\arcmin--3.3\arcmin. The maps will be significantly deeper than those planned from CMB-S4 and CCAT-P, and will have excellent fidelity to the large angular scales of our low-$z$ targets, which are difficult to probe from the ground. In combination with X-ray data from \erosita\ and \xrism\ we will transform our current static view of galaxy clusters into a full dynamic picture by measuring the internal intra-cluster medium (ICM) velocity structure with the kinematic SZE, X-ray spectroscopy, and the power spectrum of ICM fluctuations. Radio observations from ASKAP and MeerKAT will be used to better understand the connection between ICM turbulence and shocks with the relativistic plasma. Beyond the cluster boundary, we will combine thermal SZE maps from OLIMPO with X-ray imaging from eROSITA to measure the thermodynamics of the WHIM residing in filaments, providing a better understanding of its properties and its contribution to the total baryon budget.}

\maketitle

\section{Scientific Motivation}
\label{sec:motivation}

Structure formation is a dynamic, with gravity driving growth via both smooth accretion and mergers \cite{Muldrew2015}. Within the largest objects, galaxy clusters, this growth is the primary source of coherent internal motions \cite{Nelson2014}, although feedback processes such as outbursts from active galactic nuclei (AGN) also contribute \cite{McNamara2007}. Numerous signatures of these motions have been observed, including discontinuities in the ICM \cite{Markevitch2007}, the power spectrum of fluctuations in the ICM \cite{Zhuravleva2014}, and diffuse radio synchrotron from relativistic particle acceleration \cite{vanWeeren2019}. While spectroscopy of member galaxies has long been established as a minimally biased probe of the dark matter (DM) velocity \cite{Dhayaa2022}, only a handful of measurements of ICM velocity exist (e.g., \cite{Bulbul2012,Sayers2019}). As a result, many open questions still remain about both the underlying ICM physics that impacts velocities \cite{ZuHone2018}, and the overall level of motions outside of thermal equilibrium \cite{Gianfagna2023}. OLIMPO, in combination with \erosita\ and \xrism, will for the first time deliver resolved high significance measurements of the ICM velocity structure from the core region to the outskirts in order to address these questions.

Accretion occurs mainly along filaments connecting collapsed structures, with galaxy clusters typically having multiple such connections \cite{Gouin2021}. The baryons residing in these filaments are thought to comprise roughly half of the overall population, predominantly in the form of a WHIM \cite{Nicastro2018}. While new techniques, particularly dispersion measures from fast radio bursts, have emerged as promising avenues to obtain an overall WHIM census \cite{Macquart2020}, alternative approaches are still needed to understand the thermodynamic state of the WHIM. A small number of X-ray spectroscopic measurements exist, but they all indicate gas temperatures an order of magnitude higher than the predicted $\sim 0.1$~keV (e.g., \cite{Eckert2015}), likely due to selection effects and/or systematics related to probing such temperatures with existing instrumentation. By combining data from OLIMPO and \erosita\ we will measure WHIM temperatures without spectroscopy, while also accurately determining the line of sight extent of the filaments in order to mitigate possible selection biases related to filament length.

\section{The OLIMPO Instrument}
\label{sec:instrument}

OLIMPO was flown in 2018 as a differential Fourier transform spectrometer (DFTS) using kinetic inductance detectors (KIDs) to probe 130--520~GHz with 1.8~GHz resolution \cite{Paiella2019, Masi2019, Coppolecchia2020}. While a telemetry malfunction prevented scientific observations, the flight demonstrated cryogenic and detector performance in good agreement with pre-flight estimates. We aim to build upon the proven aspects of the OLIMPO instrument while making the following upgrades: remove the DFTS to create a more sensitive four-band photometer for ultra-deep SZE observations; expand the field of view (FOV) of the co-aligned focal planes to 22\arcmin\ while reducing the detector spacing to (f/\#)$\lambda$; and update the telemetry, power system, and drive motors for scanning the telescope (see Table~\ref{tab:olimpo}). To minimize risk, the KID geometry and fabrication is almost identical to the flight-proven system. We are proposing OLIMPO to NASA for a long-duration Antarctic stratospheric balloon flight in 2027.

\begin{table}[b]
    \centering
    \caption{OLIMPO Instrument Parameters}
    \label{tab:olimpo}
    \begin{tabular}{l|cccc|cccc}
    & \multicolumn{4}{|c|}{Proposed Upgrade} & \multicolumn{4}{|c}{2018 Flight - Photometric} \\
    Band (GHz) & 145 & 250 & 365 & 460 & 150 & 250 & 350 & 460 \\\hline
    Number of Detectors & 55 & 151 & 313 & 511 & 19 & 37 & 23 & 41 \\
    Detector NET ($\mu$K Hz$^{-1/2}$) & 58 & 70 & 255 & 834 & 149 & 110 & 838 & 3685 \\
    PSF FWHM (\arcmin) & 3.3 & 1.9 & 1.3 & 1.0 & 4.3 & 2.7 & 1.8 & 1.3 \\ 
    Projected Map Noise ($\mu$K-amin) & 1.2 & 0.9 & 5.2 & 2.9 & 2.7 & 2.9 & 27.9 & 255
    \end{tabular}
\end{table}

\section{Planned Observations and Ancillary Data}
\label{sec:data}

To achieve our goals of mapping the internal ICM velocity structure and probing the WHIM, we require the highest quality SZE and X-ray data possible. The recently launched \erosita\ satellite has delivered X-ray images with unprecedented sensitivity to diffuse gas in the ICM outskirts and filaments thanks to its: stable particle background at L2, large soft-band collecting area; and raster scanning to image fields much larger than its FOV \cite{Reiprich2021}. However, even with \erosita, data of sufficient quality for our study only exist for the brightest and most nearby structures at $z \simeq 0.05$. To date, ground-based SZE observations of these degree-scale objects are completely lacking due to immense challenges related to atmospheric brightness fluctuations (e.g., \cite{Bleem2015}). OLIMPO is uniquely capable among all existing and planned SZE instruments to obtain the required data: its balloon-borne platform makes atmospheric fluctuations negligible; its 2.6~m primary mirror delivers the arcminute-scale resolution required to resolve relevant features in the ICM and WHIM; and its large, co-aligned, and densely-packed focal planes allow for efficient and deep observations of degree-scale targets.

\begin{figure}
  \centering
  \includegraphics[height=0.4\textwidth]{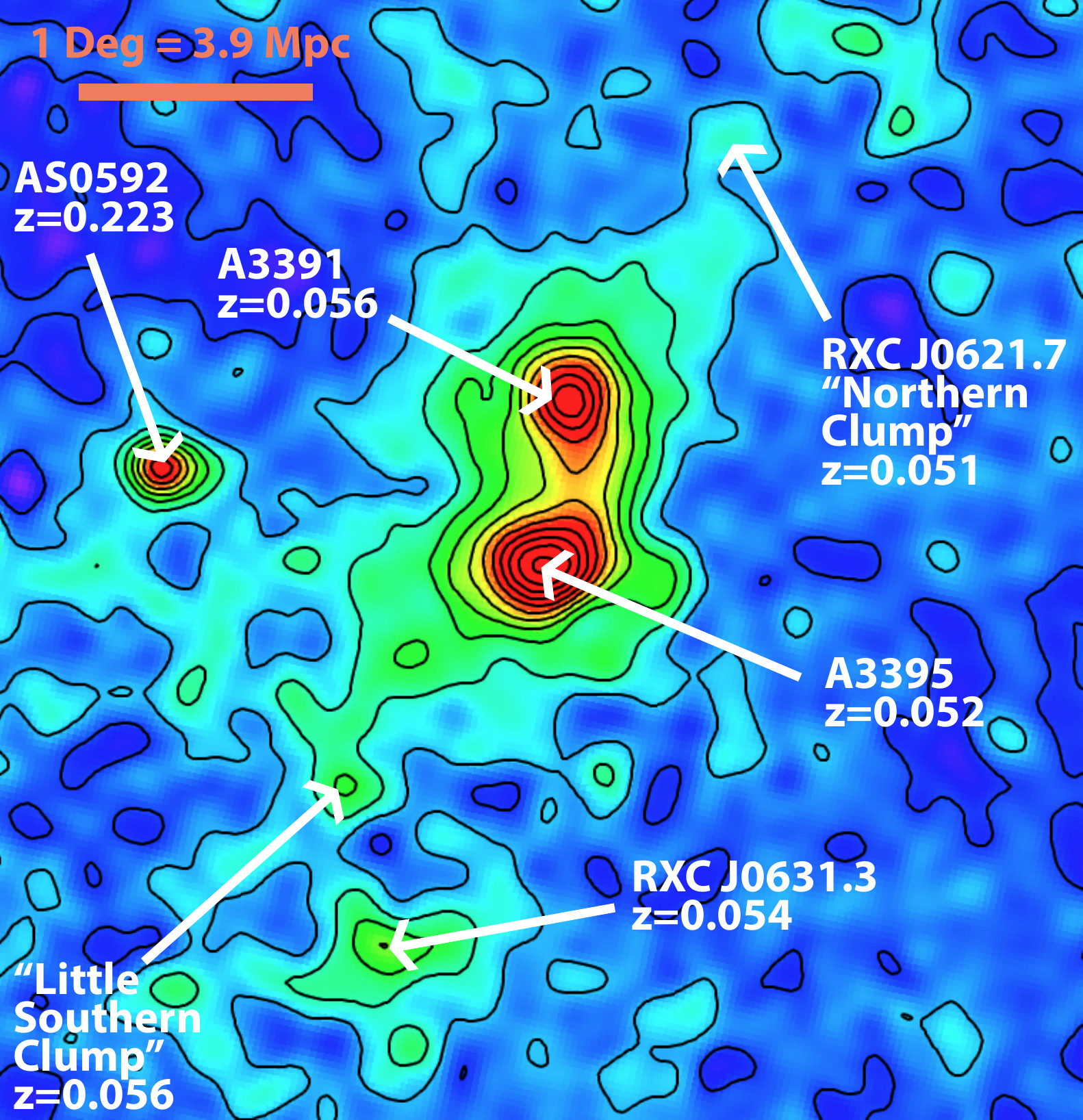}
  \hspace{0.05\textwidth}
  \includegraphics[height=0.4\textwidth]{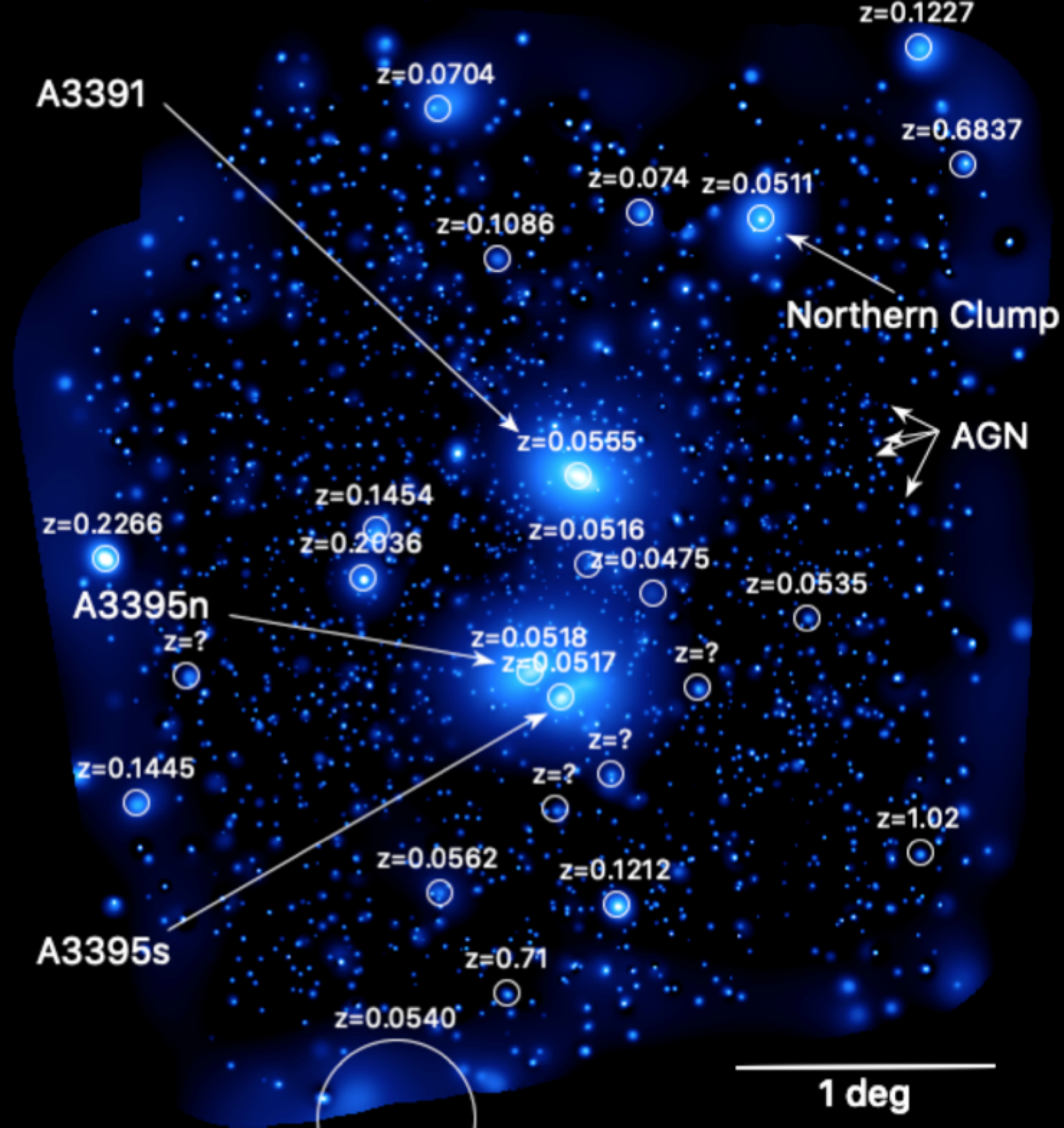}
  \caption{tSZE from \planck\ (left) and X-rays from \erosita\ (right) \cite{Planck2015, Reiprich2021}. Three bridges connect RCX J0621.7, Abell 3391, Abell 3395, and RXC J0631.3---while the bright clusters saturate the \erosita\ image, approximately 2,000, 14,000, and 3,000 photons are detected in the bridges.}
  \label{fig:A3391}
\end{figure}

\begin{table}
    \centering
    \caption{Planned OLIMPO Targets}
    \label{tab:targets}
    \begin{tabular}{l|cccc}
    Abell Number & $z$ & \planck\ SNR & \erosita\ & Notes  \\\hline
    3391/3395 & 0.05 & 17/14 & PV & x3 bridges in \planck\ \& \erosita\ \\
    3158 & 0.06 & 20 & PV & Smooth morphology w/o cool core \\
    3266 & 0.05 & 41 & PV & Merger with x2 shocks \\ 
    1644/1631 & 0.05 & 16/5 & eRASS & Long bridge, sloshing core \\
    3112 & 0.07 & 11 & eRASS & Dynamically relaxed cool-core \\
    3558/3562 & 0.05 & 19/7 & eRASS & x3 bridges in \planck\
    \end{tabular}
\end{table}

Our planned targets include those with the best \erosita\ data within the sky accessible to OLIMPO during its flight, corresponding to $0\textrm{h} \lesssim \textrm{R.A.} \lesssim 14\textrm{h}$ and $-75^{\circ} \lesssim \textrm{dec.} \lesssim -15^{\circ}$. All three deep \erosita\ performance verification (PV) fields fall within this region (see Figure~\ref{fig:A3391}), along with additional targets in the \erosita\ eRASS survey that also have deep archival \xmm\ data (see Table~\ref{tab:targets}). Complete radio coverage will be provided by the ASKAP EMU and POSSUM surveys \cite{Norris2011, Anderson2021}, along with MeerKAT for the \erosita\ PV systems \cite{Knowles2022}.

\section{Sensitivity Estimates}
\label{sec:sensitivity}

To estimate the sensitivity of OLIMPO observations, we developed a sophisticated end-to-end mock pipeline. First, we assume the measured in-flight noise equivalent powers (NEPs) and 90\% detector yield obtained from the 2018 flight \cite{Masi2019}. Based on our redesigned optical system, we then convert these NEPs to noise equivalent temperatures (NETs), corresponding to 58, 70, 255, and 834 $\mu$K~Hz$^{-1/2}$ at 145, 250, 365, and 460~GHz. We next simulate scan patterns to cover the desired target fields, including inefficiencies related to turnarounds and coverage beyond the nominal map edges. The detector performance at low frequency, in combination with our planned scan speeds and the lack of atmospheric brightness fluctuations at float altitude, suggest that the map noise power spectrum will be flat.

\begin{figure}[t]
  \centering
  \includegraphics[width=\textwidth]{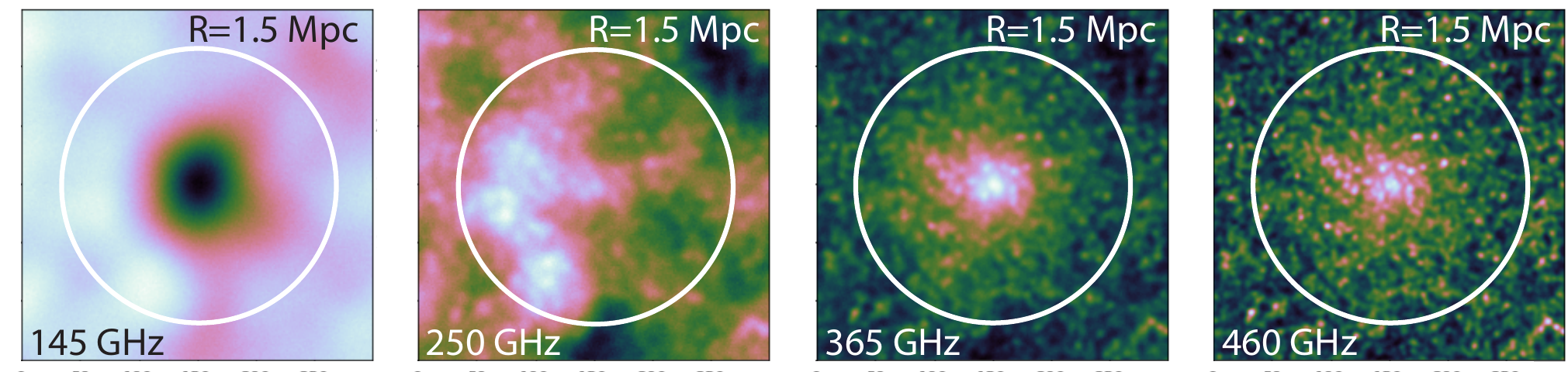}
  \caption{Mock OLIMPO observation of a representative cluster covering $1^{\circ} \times 1^{\circ}$.}
  \label{fig:mock}
\end{figure}

\begin{table}
    \centering
    \caption{Representative OLIMPO Sensitivities. For clusters we list tSZE S/N and uncertainties on the kSZE-measured velocity dispersion within 200~kpc wide azimuthally-averaged radial bins. For filaments we list per-pixel tSZE S/N and kSZE-measured velocity uncertainties within 200~kpc pixels.}
    \label{tab:sens}
    \begin{tabular}{c |c c |c c |c c}
        Abell &  \multicolumn{2}{c|}{0.4~$\le$~R~$\le$~0.6 Mpc} &  \multicolumn{2}{c|}{0.9~$\le$~R~$\le$~1.1 Mpc} &  \multicolumn{2}{c}{1.4~$\le$~R~$\le$~1.6 Mpc}\\ 
        Number  & tSZE S/N & $\sigma(v_{\textrm{disp}})$ & tSZE S/N & $\sigma(v_{\textrm{disp}})$ & tSZE S/N & $\sigma(v_{\textrm{disp}})$  \\ \hline
        3395 & 145 & $\phantom{0}$45~km/s & 115 & $\phantom{0}$60~km/s & 80 & $\phantom{0}$80~km/s \\
        3266 & 260 & $\phantom{0}$15~km/s & 125 & $\phantom{0}$50~km/s & 50 & 140~km/s \\ \hline
        3391/3395 & \multicolumn{6}{c}{tSZE S/N = 20 and $\sigma(v)$ = 45 km/s per pixel along bridge axial center} \\
        3530/3528 & \multicolumn{6}{c}{tSZE S/N = 10 and $\sigma(v)$ = 95 km/s per pixel along bridge axial center}
    \end{tabular}
\end{table}

To each noise map, we add all relevant astrophysical signals, including the thermal SZE (tSZE) and kinematic SZE (kSZE) from the target cluster and filament, along with random realizations of the galaxy population comprising the cosmic infrared background (CIB) \cite{Bethermin2017}, the primary CMB anisotropy \cite{Planck_cmb}, Galactic dust \cite{Planck_dust}, and secondary CMB anisotropy uncorrelated with the SZE from the target \cite{Reichardt2021} (see Figure~\ref{fig:mock}). The maps are then smoothed to a 3.3\arcmin\ FWHM, and a model with five parameters corresponding to the ICM optical depth, temperature, and velocity, along with the amplitude and spectral shape of the CIB, is fit to each resolution element. The parameter space is explored with Markov Chain Monte Carlo, and we aggregate samples from 40 separate mocks to account for cosmic variance. For the clusters, we assume an \erosita-measured ICM temperature accurate to 10\% as a prior. We assume a uniform map depth for all of the targets, corresponding to 1.2, 0.9, 2.3, and 5.8~$\mu$K-arcmin at 145, 250, 365, and 460~GHz. Representative SZE sensitivities based on these depths for target clusters and filaments are provided in Table~\ref{tab:sens}.

\section{Science Projections}
\label{sec:science}

\subsection{ICM Dynamics Due to Structure Formation}

\begin{figure}[t]
  \centering
  \includegraphics[height=0.37\textwidth]{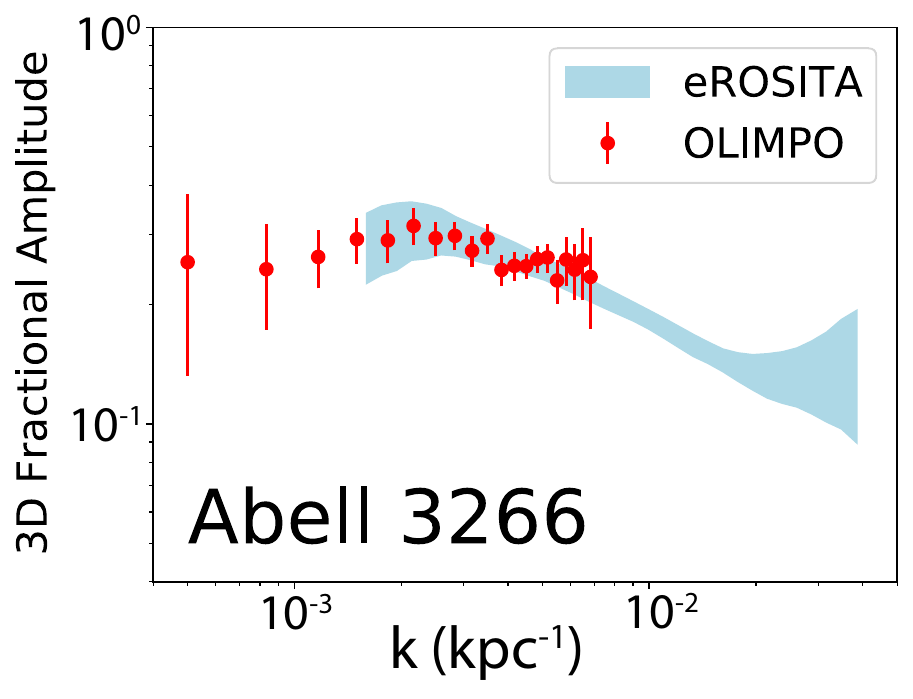}
  \hspace{0.05\textwidth}
  \includegraphics[height=0.37\textwidth]{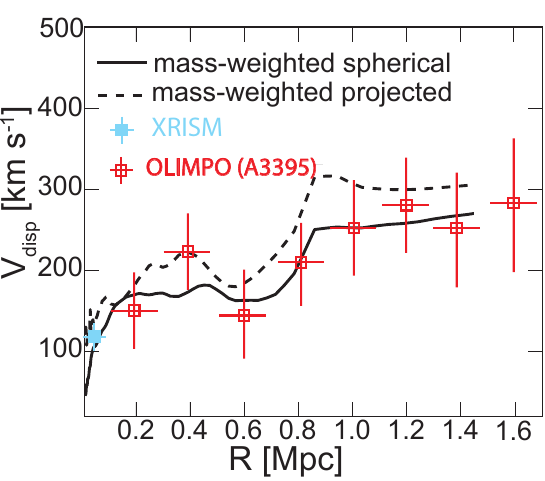}
  \caption{Left: Square root of the fluctuation power spectrum from \erosita\ for Abell 3266 (blue, \cite{Sanders2022}) and from a mock OLIMPO observation (red). Right: ICM velocity dispersion within azimuthal annuli from a mock observation of Abell 3395 compared with predictions from simulations \cite{Nagai2013}.} 
  \label{fig:power_spectrum}
\end{figure}

We plan to observe a total of 10 clusters with OLIMPO to probe, in combination with \erosita, MeerKAT/ASKAP and \xrism, the dynamics associated with structure formation in multiple complementary ways. First, we will compute the fluctuation power spectrum relative to a smooth model for the tSZE from OLIMPO and the X-ray surface brightness from \erosita\ (see Figure~\ref{fig:power_spectrum}). Via simulations, these measurements can be connected to the underlying turbulence in the ICM, probing the scale of energy injection, energy cascade, and turbulent heating rate. \erosita\ will be sensitive to scales from 0.05--0.5~Mpc, while OLIMPO will provide complementary information on scales from 0.2--2.0~Mpc.

Independent constraints on the ICM motions will be provided by direct kSZE velocity measurements from OLIMPO and X-ray microcalorimetry from \xrism. Within 200~kpc wide annular bins extending to $\simeq 1.5$~Mpc, OLIMPO will constrain the velocity dispersion to a typical precision of 50--100~km/s per cluster, sufficient to probe the expected level of motions ($\gtrsim 200$--300~km/s \cite{Nagai2013}, see Figure~\ref{fig:power_spectrum}). With \xrism, we aim to obtain more precise measurements of the velocity dispersion within the core region at R~$\le 0.1$~Mpc. We will thus obtain, for the first time, a high S/N mapping of the ICM velocity structure from the accretion-dominated outskirts to the central regions influenced primarily by AGN feedback.

By combining with radio data from MeerKAT and ASKAP, we will also be able to probe the connection between the ICM and the relativistic plasma. Turbulence is the likely source of amplified magnetic fields and particle accelerations needed to produce the radio synchrotron associated with diffuse and extended radio halos \cite{Brunetti2015}. We can directly test this model by comparing the kSZE-measured velocity structure in regions with and without halos. Merger-induced shocks produce another type of extended radio emission, relics, and there are many open questions related to these features. In particular, the Mach numbers inferred from radio measurements are generally much larger than those obtained from spatially coincident X-ray measurements of discontinuities in the ICM \cite{vanWeeren2019}. However, few comparisons exist, given the location of most relics in the outskirt regions that are difficult to probe with X-rays. OLIMPO will be able to measure weak shocks at large radii in the tSZE, for example a Mach 1.3 shock at 1~Mpc will be detected at $4\sigma$. We will thus provide a much needed independent probe of ICM shocks to help resolve the current discrepancies.

\subsection{WHIM Thermodynamics Within Filaments}

We will combine OLIMPO and \erosita\ observations of 8 bridges detected in \planck\ and/or \erosita. \erosita's unique capabilities allow it to detect $\simeq 1000$ photons from each bridge. However, with an expected temperature of $\lesssim 0.1$--0.5~keV \cite{Galarraga2021}, it is difficult to disentangle WHIM density and temperature from \erosita\ alone. By combining with tSZE images from OLIMPO, with a typical S/N of 50--100 per bridge, we expect to obtain deprojected profiles of both WHIM density and temperature relative to the bridge axis extending to at least 1~Mpc in radius with a S/N of 5 per 200~kpc wide radial bin. In addition, we will use the kSZE-measured bulk velocities of the clusters connected by the bridge, in combination with their optical spectroscopic redshifts, to determine the line of sight extent of the bridge to a precision of 0.3--1.0~Mpc. We will thus be able to search for the expected differences in WHIM properties as a function of filament length (e.g., filaments longer than 20~Mpc are predicted to be a factor of $\simeq 2$ colder than filaments shorter than 9~Mpc \cite{Galarraga2021}). Finally, OLIMPO kSZE measurements will accurately characterize the WHIM velocity structure within the filaments, which in combination with our constraints on the filament orientation relative to the line of sight will provide a detailed picture of WHIM flows within these structures.

\end{document}